\documentclass[floatfix,amssymb,superscriptaddress,
aps,prb,twocolumn]{revtex4-2}
\usepackage{stix}
\newcommand{\matrixOne}{\mathbb{1}}

\usepackage{xcolor}
\definecolor{indigo}{HTML}{4b0082}
\definecolor{firebrick}{HTML}{b22222}
\definecolor{aquamarine}{HTML}{46ad8a}
\usepackage[colorlinks=true,linkcolor=firebrick,
citecolor=aquamarine,urlcolor=indigo]{hyperref}

\usepackage{graphicx}
\usepackage{tikz}
\usepackage{amsmath}
\usepackage{amssymb}
\usepackage{mathtools}
\usepackage{braket}
\usepackage[capitalize]{cleveref}
\usepackage{makerobust}

\newcommand{\dd}{\mathrm{d}}
\newcommand{\bvec}[1]{\boldsymbol{#1}}
\newcommand{\STO}{S\lowercase{r}T\lowercase{i}O\textsubscript{3}}
\newcommand{\abinitio}{\emph{ab-initio}}
\newcommand{\angstrom}{\text{\AA}}

\DeclareMathOperator{\Tr}{\mathrm{Tr}}

\DeclarePairedDelimiter{\dbra}{\langle\hspace{-2.5pt}\langle}{|}
\DeclarePairedDelimiter{\dket}{|}{\rangle\hspace{-2.5pt}\rangle}

\renewcommand{\Im}{\mathrm{Im}}

\newcommand{\prlparagraph}[1]{\textit{#1.~---}}

\usepackage{revtex-supplement}

\begin{document}

\author{Lennart Klebl}
\affiliation{I. Institute for Theoretical Physics, Universität Hamburg,
Notkestra\ss{}e 9--11, 22607 Hamburg, Germany}
\author{Arne Schobert}
\affiliation{I. Institute for Theoretical Physics, Universität Hamburg,
Notkestra\ss{}e 9--11, 22607 Hamburg, Germany}
\author{Martin Eckstein}
\affiliation{I. Institute for Theoretical Physics, Universität Hamburg,
Notkestra\ss{}e 9--11, 22607 Hamburg, Germany}
\author{Giorgio Sangiovanni}
\affiliation{Institut für Theoretische Physik und Astrophysik and
Würzburg-Dresden Cluster of Excellence ct.qmat, Universität Würzburg, 97074
Würzburg, Germany}
\author{Alexander V.~Balatsky}
\affiliation{Department of Physics, University of Connecticut, Storrs,
Connecticut 06269, USA}
\affiliation{Nordita, Stockholm University and KTH Royal Institute of
Technology, Hannes Alfvéns väg 12, SE-106 91 Stockholm, Sweden}
\author{Tim O.~Wehling}
\affiliation{I. Institute for Theoretical Physics, Universität Hamburg,
Notkestra\ss{}e 9--11, 22607 Hamburg, Germany}
\affiliation{The Hamburg Centre for Ultrafast Imaging, 22761 Hamburg, Germany}

\title{Ultrafast pseudomagnetic fields from electron-nuclear quantum geometry}

\begin{abstract}
  Recent experiments demonstrate precise control over coherently excited
  circular phonon modes using high-intensity terahertz lasers, opening new
  pathways towards dynamical, ultrafast design of magnetism in functional
  materials. While the phonon Zeeman effect enables a theoretical description of
  phonon-induced magnetism, it lacks efficient angular momentum transfer from
  the phonon to the electron sector. In this work, we put forward a coupling
  mechanism based on electron-nuclear quantum geometry, with the inverse Faraday
  effect as a limiting case. This effect is rooted in the phase accumulation of
  the electronic wavefunction under a circular evolution of nuclear coordinates.
  An excitation pulse then induces a transient level splitting between
  electronic orbitals that carry angular momentum. First-principle simulations
  on \STO{} demonstrate that in parts of the Brillouin zone, this splitting
  between orbitals carrying angular momentum can easily reach
  $50\,\mathrm{meV}$.
\end{abstract}

\maketitle

\prlparagraph{Introduction} Coupling between angular momenta of electrons and
nuclei has been discovered more than a century ago: a body with otherwise zero
magnetization becomes magnetic when spinning---a phenomenon called Barnett
effect~\cite{barnett1915magnetization, barnett1948magnetization}. Reciprocally,
the Einstein-de-Haas effect refers to the observation that a change in
magnetization can result in mechanical rotation of an object as a
whole~\cite{richardson1908mechanical, einstein1915experimental}. These findings
were pivotal for our understanding of magnetism. More recently, experiments
demonstrated ultrafast variants of the Barnett and Einstein-de-Haas effects.
Photoexcitation can demagnetize several ferromagnets on ultrafast time
scales~\cite{kirilyuk2010ultrafast}, where circularly polarized phonons take up
angular momentum from the electronic system~\cite{dornes2019ultrafast,
tauchert2022polarized}. Reciprocally, the excitation of circularly polarized
phonons by infrared light has been demonstrated to lead to effective magnetic
fields~\cite{nova2017effective, basini2022terahertz, davies2023phononic}
allowing to switch magnetization in nanostructures~\cite{davies2023phononic} and
to the phenomenon of dynamic multiferroicity~\cite{basini2022terahertz}.

Both the Barnett and Einstein-de-Haas effects in their contemporary
understanding are focused on angular momentum transfer between nuclear- and spin
degrees of freedom. More generally, though, also the orbital electronic angular
momentum $\bvec l_\mathrm{el}$ may couple to the nuclear angular momentum $\bvec
L_\mathrm{ph}$. The time scale and effectiveness of angular momentum exchange
between the nuclear degrees of freedom and the electronic system crucially
depends on the coupling constant $K$ between the two sectors: $H = K \, \bvec
L_\mathrm{ph} \cdot \bvec l_\mathrm{el}$, which is allowed to be nonzero from
symmetry grounds ($\bvec l_\mathrm{el}$ and $\bvec L_\mathrm{ph}$ are
pseudovectors, so $H$ is a scalar). If the nuclei are driven to perform a
circular motion imposing a certain finite $\braket{\bvec L_\mathrm{ph}}$, the
coupling to the electronic system translates into an effective magnetic field
$\bvec B^\mathrm{eff} = K\,\braket{\bvec L_\mathrm{ph}}$. Clearly, the magnitude
of $K$ (or, $\bvec B^\mathrm{eff}$) is set by the microscopic mechanism giving
rise to the coupling of nuclear and electronic angular momenta in a specific
system. One mechanism to couple $\bvec l_\mathrm{el}$ and $\bvec L_\mathrm{ph}$
is the phonon Zeeman effect, i.e., genuine magnetic fields resulting from
orbital magnetic moments of the driven phonons~\cite{juraschek2017dynamical,
juraschek2019orbital, juraschek2020phonomagnetic, xiao2021adiabatically,
ren2021phonon, chaudhary2023giant}. Corresponding phonon magnetic moments are on
the order of the phonon magneton, which is typically $3-4$ orders of magnitude
smaller than the electronic one: The phonon Zeeman field is given as $B_z =
\mu_0 \mu_\mathrm{ph}/V_\mathrm{uc}$ with $\mu_\mathrm{ph}$ the phonon magnetic
moment per unit volume $V_\mathrm{uc}$. A conservative estimate via Amp\`ere's
law yields $B_z \sim 1\,\mathrm{mT}$ (see Supplementary Material (SM)~\cite{SM}
and, e.g., Ref.~\cite{juraschek2019orbital}). In contrast to the above estimate,
experiments (and theory) on $f$-electron compounds~\cite{schaack1976observation,
schaack1977magnetic, juraschek2022giant} suggest that the coupling constant $K$
between electronic and nuclear angular momenta reaches significant values, while
recent pump-probe experiments on \STO{}~\cite{basini2022terahertz} reported
phonon-induced effective magnetic fields several orders of magnitude larger than
what is expected from the phonon magneton alone.

In principle, transfer of angular momentum from the nuclear to the electronic
sector provides for an enhancement of the magnetic fields by a factor of $m_n /
m_e \sim 10^3$~\cite{geilhufe2021dynamically, juraschek2022giant}. Genuine
magnetic fields might as well be amplified by momentum space topology of the
electronic wave function~\cite{yang2013berry, yang2014giant, ren2021phonon}. In
quantum mechanics, intrinsic magnetization can further arise from the inverse
Faraday effect~\cite{pitaevskii1961electric, vanderziel1965optically,
pershan1966theoretical, kirilyuk2010ultrafast, popova2011theory,
battiato2014quantum, juraschek2020phonomagnetic}. Pseudomagnetic fields, on the
other hand, are known to exhibit enormous values for spatially inhomogeneous,
static strain fields~\cite{levy2010straininduced} and therefore pose another
candidate for effective magnetic field enhancement~\cite{merlin2024unraveling,
merlin2024magnetophononicschiralphononmisnomer}. In this Letter, we show that
driving nuclei on circular orbits can lead to pseudomagnetic fields of energy
scales dictated by the electron-phonon coupling, translating to splittings of
hundreds of $\mathrm{meV}$ in perovskite crystals. These pseudomagnetic fields
have their origin in quantum geometry of the electronic system on the manifold
of nuclear coordinates.

\begin{figure}
  \centering
  \includegraphics{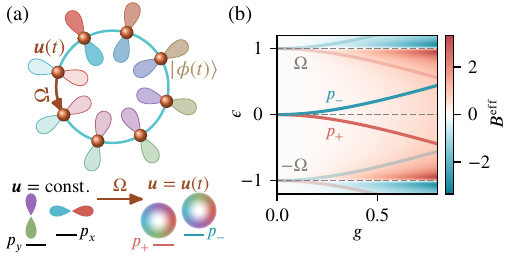}
  \caption{Jahn-Teller $E\otimes e$ model of two atomic $p$-orbitals coupled to
  two-dimensional nuclear degrees of freedom. Panel~(a) displays how the
  electronic wave function $\ket{\phi(t)}$ picks up a nontrivial phase factor
  when revolving the nuclear coordinates $\bvec u(t)$ with frequency $\Omega$.
  In addition, we highlight that the periodic $\bvec u(t)$ makes the
  orbitals split into $p_\pm$ (lower right, orbital magnetic field), whereas a constant displacement
  $\bvec u$ only leads to a splitting into real orbitals $p_{x,y}$ (lower left, crystal field).
  In panel~(b), we show numerical results for treating the Jahn-Teller $E\otimes
  e$ model in the Floquet formalism (fixing the drive frequency and amplitude
  $u=\Omega=1$). The energy levels clearly split as a function of Jahn-Teller
  coupling $g$, where red/blue lines correspond to $p_\pm$ orbital polarization.
  Floquet replica of the energy levels are included as faint red/blue lines. The
  background corresponds to the effective magnetic field
  $B^\mathrm{eff}(\epsilon)$ obtained from projecting the energy-dependent
  effective Hamiltonian $H^\mathrm{eff}(\epsilon)$ to the electronic orbital
  angular momentum operator $l_z$. We note that the size of the Zeeman splitting
  between $p_+$ and $p_-$ corresponds to the value of $B^\mathrm{eff}(\epsilon)$
  at their respective energy $\epsilon_\pm$.}
  \label{fig:pictogram}
\end{figure}

\prlparagraph{Model} To understand the origin of these pseudomagnetic fields we
consider an $E\otimes e$ Jahn-Teller model. It is the minimal description of
twofold degenerate electronic orbitals ($E$) coupling linearly to a twofold
degenerate phonon mode ($e$)~\cite{bersuker2006jahnteller}. Besides a harmonic
term for the phonon modes, the Hamiltonian includes electron-phonon coupling,
which reads
\begin{equation}
    H^\mathrm{cart} = -g\, \big( u_x \sigma_z + u_y \sigma_x \big) \,.
\end{equation}
Here, $g$ denotes the coupling constant, $\bvec u=(u_x, u_y)^T$ is the vector of
(reduced) nuclear coordinates, and $\sigma_{x,z}$ are Pauli matrices acting on
the space of electronic states $\ket{p_{x,y}}$. Written in the basis
$\ket{p_\pm}\propto\ket{p_x}\pm i\ket{p_-}$, the Hamiltonian becomes
\begin{equation}
    H = -g\,\bvec \sigma \cdot \bvec u\,,
    \label{eq:jt-model}
\end{equation}
with $\bvec \sigma = (\sigma_x,\sigma_y)^T$. If we now assume periodically
driven nuclear coordinates, e.g., $\bvec u = u\,(\cos\Omega t, \sin\Omega t)^T$,
we can treat the system in Floquet space. Following
Refs.~\cite{rodriguez-vega2018floquet, oka2019floquet, vogl2020effective}, we
construct an effective Hamiltonian (in high-frequency or low-amplitude
approximation) as
\begin{equation}
  H^\mathrm{eff} = h_0 +
  h_1 \frac1{h_0 - \Omega} h_1^\dagger +
  h_1^\dagger \frac1{h_0 + \Omega} h_1\,,
  \label{eq:heff}
\end{equation}
where the Floquet hopping $h_n$ is defined as
\begin{equation}
  \label{eq:floqelem}
  h_n = \frac1{T}\int_0^T\!\!\dd t\,e^{-in\Omega t} H\big(\bvec u(t)\big) \,,
\end{equation}
with $T = 2\pi/\Omega$ the period of the drive (for details on the Floquet and
effective Hamiltonians see SM~\cite{SM}). Note that all terms except $h_{\pm1}$
vanish due to the linear nature of $H$. Since we transformed to the basis of
$\ket{p_\pm}$ states, the terms proportional to $\sigma_z$ correspond to an
effective magnetic field:
\begin{equation}
  B^\mathrm{eff} = \frac{\Tr(H^\mathrm{eff} \sigma_z)}2 = -\frac{g^2u^2}{\Omega}\,.
  \label{eq:beff-sph}
\end{equation}
Beyond first order in $1/\Omega$, we numerically diagonalize the Floquet
Hamiltonian and obtain a quasienergy spectrum that is presented in
\cref{fig:pictogram}~(b). It is clearly visible that the splitting between the
$p_+$ (red) and $p_-$ (blue) level grows quadratically with coupling strength
even beyond the antiadiabatic limit $gu \ll \Omega$ [we set $\Omega=u=1$ in
\cref{fig:pictogram}~(b)].

Floquet perturbation theory in the $\Omega\to\infty$ limit allows us to
construct the eigenstates of the Floquet
Hamiltonian~\cite{rodriguez-vega2018floquet}. Within this limit, we complement
the above picture of an effective Hamiltonian with the quantum geometric one.
The non-adiabatic generalization of Berry's phase for a state
$\ket{\phi_\alpha(t)}$, the Aharonov-Anandan phase~\cite{aharonov1987phase,
oka2009photovoltaic, oka2019floquet}, is to leading order in $1/\Omega$ given as
\begin{equation}
  \gamma^\mathrm{AA}_\pm = \int_0^T\!\!\dd t\, \bra{\phi_\pm(t)} i
  \partial_t \ket{\phi_\pm(t)} = \pm\frac{2\pi g^2u^2}{\Omega^2}
\end{equation}
in the $p_\pm$ basis (see SM~\cite{SM}). It enters the Floquet quasienergies as
$\gamma^\mathrm{AA}_\pm/T$ and therefore generates the same $\Omega\to\infty$
result as the consideration through $H^\mathrm{eff}$:
\begin{equation}
  \frac{\gamma^\mathrm{AA}_-}{T} = \frac{-g^2u^2}{\Omega} = B^\mathrm{eff} \,,
\end{equation}%
therefore manifesting dynamical quantum geometry (see \cref{fig:pictogram}~(a)
for an illustration) as the origin of the effective magnetic field in the
$E\otimes e$ Jahn-Teller model. Even for weak coupling strength
$gu=5\,\mathrm{meV}$ and pump frequencies in the regime
$\Omega=20\,\mathrm{meV}\approx5\,\mathrm{THz}$, the effective magnetic field
reaches $B^\mathrm{eff}=1.25\,\mathrm{meV} \approx \mu_B\,22\,\mathrm{T}$.
Notably, $\gamma^\mathrm{AA}$ relates to $B^\mathrm{eff}$ ($H^\mathrm{eff}$) in
other models as well---in particular, we demonstrate that the phonon inverse
Faraday effect~\cite{shabala2024phononinversefaradayeffect} can be understood as
high-frequency limit of the Aharonov-Anandan phase (see SM~\cite{SM}). We note
that the quantum geometric picture is more general than the inverse Faraday
effect picture, as it does not require an intermediate state (for Raman
processes).

In addition to the Floquet spectrum, \cref{fig:pictogram}~(b) displays the
numerical calculation of $B^\mathrm{eff}$ in the (molecular) Jahn-Teller
$E\otimes e$ model \cref{eq:jt-model}. We obtain the energy dependency of the
effective magnetic field from projecting the energy-dependent effective
Hamiltonian to the angular momentum operator $l_z = \sigma_z$. This formulation
extends \cref{eq:heff} in that it takes all orders of $1/\Omega$ into account
via inversion of the Floquet Green's function's $n=0$ block
(cf.~\cite{vogl2020effective} and SM~\cite{SM}), i.e., $H^\mathrm{eff}(\epsilon)
= (\mathcal G(\epsilon)_{0,0})^{-1}$. As we set $\Omega = u = 1$ in
\cref{fig:pictogram}~(b), the coupling strength $g$ acts as an energy scale. We
highlight that $B^\mathrm{eff}$ reaches values on the order of $g$ and $\Omega$
as soon as $g$ becomes sizable.

\begin{figure}
  \centering
  \includegraphics{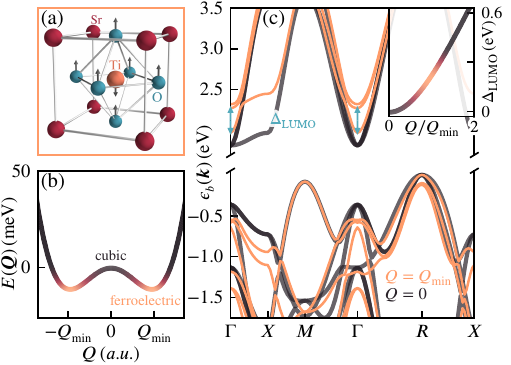}
  \caption{Electron-phonon coupling of the ferroelectric mode in \STO. Panel~(a)
  displays the crystal structure in the cubic phase and additionally highlights
  the displacements of Ti and O atoms in the (unstable) ferroelectric phase
  (gray arrows). In panel~(b) we plot the energy $E(\bvec Q)$ of the
  electronic system as a function of the ferroelectric mode amplitude $Q$.
  Panel~(c) presents the \abinitio{} electronic structure of \STO{} in the cubic
  phase (thick black lines, $Q=0$) and the ferroelectric phase (orange lines,
  $Q=Q_\mathrm{min}$). The inset shows the behavior of
  the valence band gap $\Delta_\mathrm{LUMO}$ as a function of the ferroelectric
  mode amplitude $Q/Q_\mathrm{min}$.}
  \label{fig:sto}
\end{figure}
\prlparagraph{Material Realization}
While sizable linear electron-phonon coupling acting on degenerate electronic- as well as phononic modes might seem like a frequently encountered situation in materials, finding systems where the nuclei can be driven by external fields in a way that they rotate at significant amplitudes is expected to be more difficult.
In Buckyball fullerenes, Jahn-Teller physics is known to
play an important role~\cite{obrien1993jahn, bersuker2006jahnteller}. With
additional atoms trapped in the carbon cages, so-called
endofullerenes~\cite{tellgmann1996endohedral, gromov2002optical, umran2015study,
krachmalnicoff2016dipolar} can host polar phonon modes. Another recent material
example that hosts soft, polar phonons is \STO{}: The $T_{1u}$ modes responsible
for the material's quantum paraelectric behavior are known to be controllable
with terahertz radiation~\cite{li2019terahertz, kozina2019terahertzdriven,
basini2022terahertz, yang2023terahertz} and couple strongly to the $T_{2g}$
electron sector. \Cref{fig:sto}~(a) displays the crystal structure of \STO{},
including an illustration of the ferroelectric displacement, where the titanium
atoms move opposite to their surrounding oxygen octahedra and create a net
polarization. At the ferroelectric minimum $Q_\mathrm{min}$, the titanium atoms
are displaced by roughly $0.11\,\text{\AA}$ relative to the oxygen octahedra. We
further show that the Higgs-like potential for the ferroelectric mode is shallow
[see panel~(b)], underlining that amplitudes on the order of the ferroelectric
minimum $Q_\mathrm{min}$ can be reached by terahertz
driving~\cite{kozina2019terahertzdriven, basini2022terahertz}. In order to
analyze the effect of the ferroelectric mode on the electronic sector, we
calculate the \abinitio{} band structure of \STO{} [cf.~\cref{fig:sto}~(c)] in
both its symmetric coordination (black) as well as under the influence of a
polar (ferroelectric) displacement (orange). A significant gap
$\Delta_\mathrm{LUMO} \approx 200\,\mathrm{meV}$ at $\bvec k=\Gamma$ opens when
the atoms are displaced to the ferroelectric minimum $Q_\mathrm{min}$ [see inset
of \cref{fig:sto}~(c)].

\begin{figure}%
  \centering
  \includegraphics{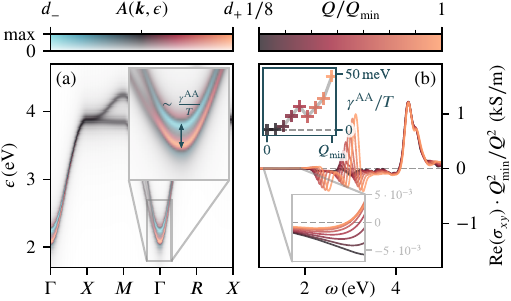}
  \caption{Floquet spectral function~(a) and real part of the Hall component $\sigma_{xy}$ of the optical conductivity~(b) for the
  \abinitio{} model of \STO. We assume a drive frequency of
  $\Omega=30\,\mathrm{meV}\approx7\,\mathrm{THz}$. In panel~(a), we set the
  displacement to $Q=Q_\mathrm{min}$, i.e., the ferroelectric minimum. We note
  that the splitting at $\bvec k=\Gamma$ is of slightly smaller size than the static gap
  $\Delta_\mathrm{LUMO}$ [cf.~\cref{fig:sto}~(c)]. The color encodes orbital
  polarization with respect to the Ti $d_\pm$ orbitals, and the brightness
  corresponds to the overall spectral weight in the $d_\pm$ sector. The inset
  verifies that the bands close to $\Gamma$ are split by an effective orbital
  magnetic field, i.e., strongly polarized in terms of  $d_+$- (red) and
  $d_-$-orbitals (blue). Panel~(b) displays the real part of the Hall conductivity $\mathrm{Re}(\sigma_{xy})$ for varying displacement $Q$. We rescale $\mathrm{Re}(\sigma_{xy})$ by the
  quadratic ratio $Q^2/Q_\mathrm{min}^2$ and thereby demonstrate
  $\sigma_{xy}\propto Q^2$ for excitations into the condcution band sector
  $\omega\gtrsim3\,\mathrm{eV}$. The upper inset displays the evolution of the
  Aharonov-Anandan phase difference between Ti-$d_\pm$ orbitals at the Brillouin
  zone center $\bvec k=\Gamma$. Its quadratic scaling for small $Q$ translates
  into quadratic scaling of $\sigma_{xy}$ for small $Q$, as seen in the lower inset for the in-gap behavior of $\mathrm{Re}(\sigma_{xy})$.}
  \label{fig:fullmodel}
\end{figure}

In order to extend our analysis to the driven setting, we apply the newly
developed downfolding procedure outlined in Ref.~\cite{schobert2024abinitio} to
arrive at an \emph{ab-initio} Wannier model describing oxygen-$p$ and
titanium-$T_{2g}$ orbitals coupled to a lattice-spring model for the phonons
(see SM~\cite{SM} for details). The time dependent electronic Hamiltonian is
therefore given by
\begin{equation}
    \hat H(\bvec k,t) = \hat H_\mathrm{W}(\bvec k) + \sum_i Q \hat g_i(\bvec k) \, \big( u_{i,x} \cos(\Omega t) + u_{i,y} \sin(\Omega t) \big)\,,
\end{equation}
where $\hat H_\mathrm{W}(\bvec k)$ is the Wannier Hamiltonian, $\hat g_i(\bvec
k)$ the electron-phonon coupling matrix (at phonon momentum $\bvec q=0$) for a
given atom $i$ (where $\sum_i$ runs over all atoms in the unit cell), and
$u_{i,x/y}$ are two phonon eigenvectors from the $T_{1u}$ sector. Using
\cref{eq:floqelem}, we can read off the Floquet hoppings as
\begin{equation}
    \label{eq:wannierfloq}
    h_0 = \hat H_W(\bvec k) \,, \quad h_{\pm1} = \frac Q2\hat g_i(\bvec k)\big( u_{i,x} \pm i u_{i,y} \big)\,.
\end{equation}
With the above Floquet hopping matrices at hand, we set up the Floquet
Hamiltonian $H_F(\bvec k)$ (see SM~\cite{SM} for details) and diagonalize it.
Due to the multitude of bands and Floquet replica, we choose to analyze the
model in terms of its spectral function
\begin{equation}
    A_\phi(\bvec k,\epsilon) = \Im \braket{\phi|\mathcal G(\bvec k,\epsilon-i\eta)_{0,0}|\phi} \,,
\end{equation}
obtained from the $n=0$ sector of the Floquet Green's function $\mathcal G(z)$.
As most of the conduction band weight is in the titanium-$T_{2g}$ sector, we use
complex orbitals $d_\pm \propto d_{xz} \pm i d_{yz}$ for the projected state
$\ket{\phi}$. \Cref{fig:fullmodel}~(a) shows how \STO{} responds to dynamically
driving the $T_{1u}$ phonon mode: The conduction band splits into two bands that
display orbital polarization in the $d_\pm$ basis. Notably, the splitting is
pronounced only around $\Gamma$---by moving away from the zone center we quickly
recover the unperturbed conduction band structure [cf. \cref{fig:sto}~(c)]. The
inset of \cref{fig:fullmodel}~(b) displays the $\Gamma$-point AA-phase
difference between $d_+$ and $d_-$ orbitals as a function of ferroelectric
amplitude $Q$, clearly demonstrating that the splitting can be attributed to
quantum geometric effects.
Estimating the momentum-local effective magnetic field at $\Gamma$ through the
gap size yields $B^\mathrm{eff}(\bvec k=\Gamma)\approx 0.07\,\mathrm{eV}\approx
\mu_B\,1.2\,\mathrm{kT}$. We stress that a local magnetic field in real space
corresponds to a momentum average and due to the lack of a splitting away from
$\Gamma$ results in a significantly lower value.

In addition to tracing the pseudomagnetic field generated by the rotating
ferroelectric mode through orbital polarization in the spectral function, we
calculate the $n=0$ Floquet optical conductivity $\sigma_{ij}(\omega) \equiv
\sigma_{ij}^{(0)}(\omega)$ (homodyne component, i.e., ``on-shell"{} response at
the probe frequency, see SM~\cite{SM} for details)~\cite{kumar2020linear,
cupo2023optical, shah2023magnetooptical}. In the calculation of $\sigma^{(0)}$,
we assume equilibrium occupations given that the band gap $\Delta$ of \STO{} is
by far the largest energy scale ($\Delta \gg \Omega$ and $\Delta \gg gQ$). We
obtain an optical conductivity with pronounced resonance structure and an
increase of both $\sigma_{xy}$ and $\sigma_{xx}$ upon approaching the absorption
edge $\Delta \approx 2.3\,\mathrm{eV}$.
The band gap is known to be underestimated in semi-local density functional theory, which is why we discuss $\omega=1\,\mathrm{eV}$ as a frequency inside the gap in the following.
For probe frequencies inside the optical
gap, only the anti-hermitian, i.e., non-dissipative, components of $\sigma_{ij}$
remain nonzero. \Cref{fig:fullmodel}~(b) demonstrates that the anti-hermitian
(real, non-dissipative) part of the Hall component behaves like
$\mathrm{Re}(\sigma_{xy})\sim Q^2$ for $Q \ll Q_\mathrm{min}$, with
$Q_\mathrm{min}$ the driving amplitude at the ferroelectric minimum
[cf.~\cref{fig:sto}~(b)]. This component ($\mathrm{Re}(\sigma_{xy})$, see lower
inset for in-gap behavior) leads to a finite in-gap Faraday rotation
$\theta_F$~\cite{freiser1968survey, suits1972faraday, brynildsen2013verdet}. The
estimated value of $\theta_F$ is in line with the experimental results by
Basini~\emph{et.al.}~\cite{basini2022terahertz}: We obtain $\theta_F \cdot
\lambda/d \lesssim 100\,\mathrm{\mu rad}$ at $\omega \approx 1\,\mathrm{eV}$,
where $d$ is the penetration depth of the $\mathrm{THz}$ pulse and $\lambda$ the
wavelength of the optical probe (see SM~\cite{SM}). In
Ref.~\cite{basini2022terahertz}, the Faraday rotation reaches up to roughly
$200\,\mathrm{\mu rad}$ (see Extended Data Fig.~6 therein), while the
penetration depth of the pump pulse is estimated as $d\approx 4\lambda$, which
therefore is in good agreement with our theory. We note that for a semi-infinite
system, the real part of the Kerr angle is zero inside the optical
gap~\cite{you1998generalized, mansuripur2000magneto, protopopov2014practical}.
So the report of finite Kerr rotation (which is of similar magnitude as the
Faraday rotation) in Ref.~\cite{basini2022terahertz} might be due to finite
penetration depth of the pump pulse including Fabry-Perot-like repeated
scattering at the top and emergent bottom surfaces~\cite{you1998generalized,
tse2010giant, catarina2020magneto}.


\prlparagraph{Summary and discussion} We propose a mechanism to
obtain giant pseudomagnetic fields for electrons in both molecules and crystals.
These fields are generated by the Aharonov-Anandan phase, a non-adiabatic
extension to the concept of Berry's phase, which the electron wave function
collects through evolution along the trajectories of rotating nuclear degrees of
freedom.

Through the example of an $E\otimes e$ Jahn-Teller model, we illustrate that the
pseudomagnetic field strength is proportional to the squared driving amplitude
$g^2u^2$, as well as the inverse pump frequency $1/\Omega$ in the high-frequency
limit. Even in this limit, fields in the range of tens of Teslas are in reach
for terahertz pump pulses. We further demonstrate that upon treating the
perovskite crystal \STO{} with \abinitio{} techniques, the resulting
pseudomagnetic field can become huge ($\sim\mathrm{kT}$) when considered as an
energy gap between chiral orbitals at special points in momentum space (for pump
frequencies in the intermediate regime, $\Omega\approx7\,\mathrm{THz}$). The
local pseudomagnetic field (i.e., a momentum-integrated version) is expected to
be reduced considerably.

By means of the Floquet optical conductivity, we were able to estimate the Faraday
rotation in \STO{} for pump-probe experiments. It is therefore tempting to
compare our results to the experimental reports of dynamical
multiferroicity~\cite{juraschek2017dynamical} in \STO{} by
Basini~\emph{et.al.}~\cite{basini2022terahertz}. The authors interpret the
measured Faraday rotation in terms of effective magnetization
$\sim0.16\,\mu_B\mathrm{T}/\mathrm{u.c.}$, but the underlying observable is the
Faraday rotation $\theta_F\sim0.1\,\mathrm{mrad}$. This value for the Faraday rotation
is confirmed by our result for probe frequencies inside the band gap, and even
surpassed for large amplitudes. This is plausible for the following reason: The
phonon mode is expected to be delocalized due to quantum nuclear effects, i.e.,
the average amplitude might only be a fraction of the ferroelectric minimum
$Q_\mathrm{min}$ and also the phase will not be sharply defined.

According to \cref{eq:beff-sph} and the Faraday rotation results for the
\abinitio{} model, we expect a quadratic dependence of $B^\mathrm{eff}$ on the
driving electric field under the assumption of classical nuclei and linear
electron-nuclear coupling, which is in accord with the observations of the
dependence observed by Basini~\emph{et.al.}~\cite{basini2022terahertz}. Our
findings strongly suggest that in \STO{}, electron-nuclear quantum geometry can
be seen as the source of dynamical multiferroicity~\cite{juraschek2017dynamical,
li2019terahertz,basini2022terahertz, zhuang2023lightdriven}.
Apart from perovsikes close to ferroelectric
phases~\cite{rowley2014ferroelectric, basini2022terahertz}, we generically
expect similar pseudomagnetic fields from electron-nuclear quantum geometry in
materials with significant electron-phonon coupling of (polar) phonon modes to
degenerate electronic levels, such as
endofullerenes~\cite{tellgmann1996endohedral, gromov2002optical, umran2015study,
krachmalnicoff2016dipolar, obrien1993jahn, bersuker2006jahnteller}. Furthermore,
we expect nonlinear Jahn-Teller (e.g., $E\otimes e$) molecules to show this type
of efficient angular momentum transfer from the vibrational to the electronic
sector. Notably, the pseudomagnetic fields can be switched on ultrafast
picosecond time scales.

\prlparagraph{Note added} During the final preparation of this manuscript, the
following works discussing orbital magnetic fields in driven \STO{}
appeared:~\cite{urazhdin2024orbitalmomentgenerationcircularly,
libbi2024ultrafastquantumdynamicsmathbfmathrmsrtio3}.

\bigskip

\begin{acknowledgments}
  We thank S.~Bonetti,  M.~Bunney, A.~Fischer, M. Geilhufe and N.~Spaldin for
  fruitful discussions. We gratefully acknowledge support from the Deutsche
  Forschungsgemeinschaft (DFG, German Research Foundation) through FOR 5249
  (QUAST, Project No. 449872909, TP5), EXC 2056 (Cluster of Excellence ``CUI:
  Advanced Imaging of Matter'', Project No. 390715994) and the Würzburg-Dresden
  Cluster of Excellence on Complexity and Topology in Quantum Matter ct.qmat
  (EXC 2147, Project ID 390858490). A.B was supported by European Research
  Council under the European Union Seventh Framework ERS-2018-SYG 810451 HERO,
  and the University of Connecticut. We further acknowledge computational
  resources provided by the North-German Supercomputing Alliance (HLRN) as well
  as computational resources provided through the JARA Vergabegremium on the
  JARA Partition part of the supercomputer JURECA~\cite{krause2018jureca} at
  Forschungszentrum Jülich.
\end{acknowledgments}

\bibliography{references}

\supplement{Supplementary Information:\\
Ultrafast pseudomagnetic fields from electron-nuclear quantum geometry}

\section{Circularly polarized phonons and Amp\`ere's law}
We estimate an upper limit for the strength of magnetic fields generated by
circularly polarized phonons by considering them as classically rotating
charges. The magnetic moment of a single charge $q$ revolving at radius $r$ with
frequency $\Omega$ is given as
\begin{equation}
  \mu_\mathrm{ph} = I \, A = q\Omega \, \pi r^2 \,,
\end{equation}
Assuming extreme conditions $q=2e$, $r=0.1\,\mathrm{\angstrom}$ and
$\Omega/2\pi=6\,\mathrm{THz}$ as well as a very small unit cell of
$V_\mathrm{uc} = 5\,\mathrm{\angstrom^3}$, we obtain
\begin{equation}
  B_z = \mu_0 \, \frac{ \mu_\mathrm{ph} }{V_\mathrm{uc}} \approx
  1\,\mathrm{mT}\,.
\end{equation}

\section{Floquet perturbation theory of the $E\otimes e$ Jahn-Teller model}
\label{app:floqexe}
The full Hamiltonian of the $E\otimes e$ Jahn-Teller model reads
\begin{equation}
    H^\mathrm{JT} = \frac{\bvec u^2 - \partial_{\bvec u}^2}{2} + g\begin{pmatrix}
        -u_x & -u_y \\
        -u_y & u_x
    \end{pmatrix} = \frac{\bvec u^2 - \partial_{\bvec u}^2}{2} + H^\mathrm{cart}\,,
\end{equation}
where $\bvec u = (u_x, u_y)^T$ is the nuclear coordinate and the matrix acts on
electronic (e.g.~$\ket{p_{x,y}}$) states. Our work focuses on the form of the
coupling rather than on the kinetics of the nuclei, so we transform to the
complex basis $\ket{p_\pm}$:
\begin{equation}
    \ket{p_\pm} = \frac1{\sqrt2} \, \big( \ket{p_+} \pm i \ket{p_-} \big)\,, \qquad
    U = \frac1{\sqrt2}\begin{pmatrix}
        1 & -i \\
        1 & i
    \end{pmatrix}\,, \qquad H = U H^\mathrm{cart} U^\dagger = -g\begin{pmatrix}
        0 & u_x - iu_y \\
        u_x + iu_y & 0
    \end{pmatrix} = -g \bvec \sigma \cdot \bvec u\,,
\end{equation}
with $\bvec \sigma = (\sigma_x, \sigma_y)^T$. We assume a periodic evolution of
nuclear coordinates, and therefore set $\bvec u = u\,(\cos\Omega t, \sin\Omega
t)^T$. The time dependent Hamiltonian becomes
\begin{equation}
    H(t) = -gu\begin{pmatrix}
        0 & e^{-i\Omega t} \\
        e^{i\Omega t} & 0
    \end{pmatrix} \,.
\end{equation}
In Floquet space, we determine the ``hoppings'' as
\begin{equation}
    h_n = \frac1T\int_0^T\!\!\dd t\,e^{-in\Omega t}\,H(t) = -gu\begin{cases}
        \rho_\pm &\text{for}~n=\pm 1 \\
        0 &\text{else}
    \end{cases} \,,\quad\text{with}~\rho_+ = \begin{pmatrix}
        0 & 1 \\ 0 & 0
    \end{pmatrix}~\text{and}~\rho_- = \begin{pmatrix}
        0 & 0 \\ 1 & 0
    \end{pmatrix}\,.
\end{equation}
The Floquet Hamiltonian can be represented in matrix form as
\begin{equation}
    \label{eq:supp:hfmat}
    H_F = \begin{pmatrix}
        \ddots & h_1 \\
        h_1^\dagger & h_0+2\Omega & h_1 \\
        & h_1^\dagger & h_0+\Omega & h_1  \\
        & & h_1^\dagger & h_0 & h_1  \\
        & & & h_1^\dagger & h_0-\Omega & h_1  \\
        & & & & h_1^\dagger & h_0-2\Omega & h_1  \\
        & & & & & h_1^\dagger & \ddots \\
    \end{pmatrix}\,.
\end{equation}
Following Ref.~\cite{rodriguez-vega2018floquet}, we perform high-frequency
perturabtion theory for the $\ket{p_\pm}$ states with the result
\begin{equation}
    \ket{p_\pm^{(1)}(t)} = \sum_{n\neq0} \frac{h_{-n}\,e^{in\Omega t}}{n\Omega}
    \ket{p_\pm} = \bigg( \frac{h_{1}e^{-i\Omega t}}{-\Omega} +
    \frac{h_{-1}e^{i\Omega t}}{\Omega} \bigg)\ket{p_\pm}\,.
\end{equation}
From this expression, we can calculate the Aharonov-Anandan
phase~\cite{aharonov1987phase, oka2009photovoltaic, oka2019floquet} as
\begin{equation}
    \frac{\gamma^\mathrm{AA}_{\pm}}T = \frac1T\int_0^T\dd t
    \braket{p_\pm^{(1)}(t) | i\partial_t | p_\pm^{(1)}(t)} =
    \bra{p_\pm}\bigg(\frac1{\Omega}\big[h_1^\dagger,h_1\big]\bigg) \ket{p_\pm}
    = \mp\frac{-g^2u^2}{\Omega} \,.
\end{equation}
As we shall see below, it perfectly coincides with the magnetic fields generated
in the effective Hamiltonian picture.

\section{Floquet Green's functions and effective Hamiltonians}
\label{sec:supp:floqg}
The resolvent operator in Floquet space can be defined as the matrix inverse of
$H_F$ [cf.~\cref{eq:supp:hfmat}]:
\begin{equation}
    \mathcal G(\omega) = \frac1{\omega - H_F} \,.
\end{equation}
We define the effective Hamiltonian as the re-inversion of the $n=0$ Floquet
block of $\mathcal G$, i.e.,
\begin{equation}
    H^\mathrm{eff}(\omega) = \big( \mathcal G_{0,0}(\omega) \big) ^{-1} \,.
\end{equation}
This creates, in addition to the time-averaged Hamiltonian $h_0$, self-energy
like terms that, upon inversion, yield the correct propagator. In order to
circumvent points where $\mathcal G_{0,0}$ is not invertible, we add an
imaginary broadening to $\omega$, i.e., $\omega\leftarrow\omega + i\eta$. To
leading order in $1/\Omega$ and for $\omega=0$, we can express $H^\mathrm{eff}$
as~\cite{rodriguez-vega2018floquet, oka2019floquet, vogl2020effective}
\begin{equation}
    H^\mathrm{eff} = h_0 + h_{-1}\,\frac1{h_0-\Omega}\,h_{-1}^\dagger +
    h_{-1}^\dagger\,\frac1{h_0+\Omega}\,h_{-1} \,,
    \label{eq:supp:heff-small}
\end{equation}
i.e., propagating to the next Floquet block and back. For the $E\otimes e$
Jahn-Teller model, the construction of $H^\mathrm{eff}$ becomes particularly
simple since $h_0=0$:
\begin{equation}
    H^\mathrm{eff} = h_{-1} \frac1{-\Omega} h_{-1}^\dagger + h_{-1}^\dagger \frac1\Omega h_{-1} = -\frac{g^2u^2}{\Omega} \sigma_z \,.
\end{equation}
Due to our choice of basis as $\ket{p_\pm}$, the effective Hamiltonian
corresponds to a pseudomagnetic field $B^\mathrm{eff} =
\Tr(H^\mathrm{eff}\sigma_z)/2 = -g^2u^2/\Omega$.


\section{\abinitio{} parameters for \STO{}}
\label{app:computationalparameters}

\begin{figure}
    \centering
    \includegraphics{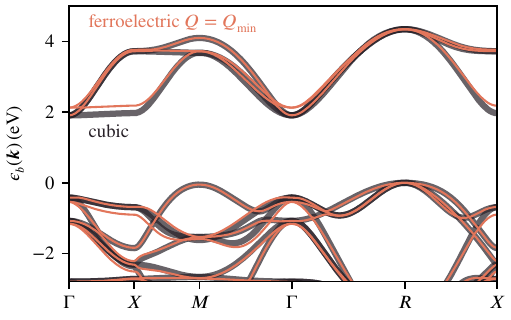}
    \caption{Band structure of the Wannier model (``Model III'') in the cubic
    phase (black) and the ferroelectric phase (orange). The conduction bands are
    gapped with $\Delta_\mathrm{LUMO}\approx 0.22\,\mathrm{eV}$ at $\bvec
    k=\Gamma$ when the titanium atoms are shifted by $\sim0.11\,\text{\AA}$
    relative to the oxygen octahedra.}
    \label{fig:supp:model3}
\end{figure}

All Density Functional (Perturbation) Theory calculations are carried out using
\textsc{Quantum ESPRESSO}~\cite{Giannozzi2009,Giannozzi2017}. For the
transformation of the electronic energies and electron-phonon couplings to the
Wannier basis, we use \textsc{Wannier90}~\cite{Mostofi2014} and the \textsc{EPW}
code~\cite{Noffsinger2010,Ponce2016}. We have applied the Perdew-Burke-Ernzerhof
(PBE) functional~\cite{Perdew1996} in combination with Ultrasoft
pseudopotentials~\cite{Vanderbilt1990} from the SSSP
library~\cite{Prandini2018,Prandini2021}. Furthermore, we used a lattice
constant of 3.936~\AA, set the plane-wave cutoff to 120~Ry and used
$4\times4\times4$ points for both the $\bvec k$ and the $\bvec q$ meshes.

\begin{figure}
    \centering
    \includegraphics{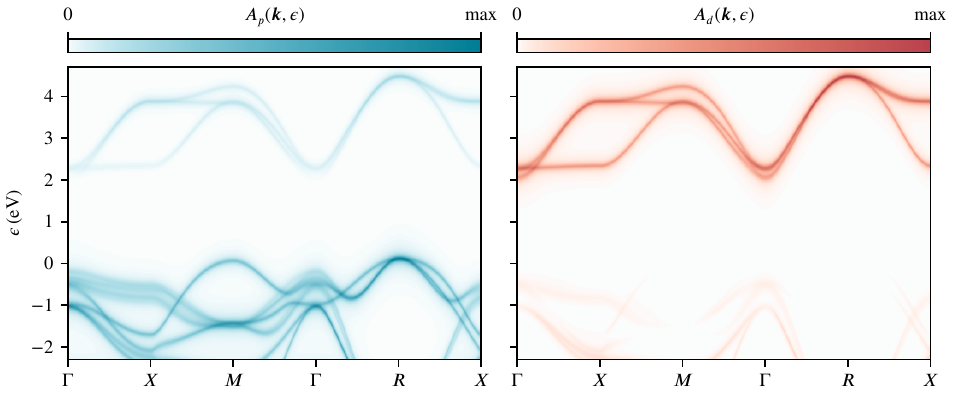}
    \caption{Floquet spectral function of \STO{} at $Q=Q_\mathrm{min}$ driven
    with a frequency $\Omega=30\,\mathrm{meV}$. The left plot shows the
    projection of the spectral function to the oxygen-$p$ orbitals, and the
    right plot the projection to the titanium-$T_{2g}$ block.}
    \label{fig:supp:spectral}
\end{figure}

The downfolded model, which is used to calculate the Floquet optical
conductivity and the Floquet spectral function of \STO{}, is described in detail
as ``Model III'' in Ref.~\cite{schobert2024abinitio}. The model is based on a
linearized electron-nuclear Hamiltonian
\begin{equation}
\hat H(\bvec k, \bvec Q) = \hat H_W(\bvec k) + \sum_{i,\alpha} \hat{g}_i(\bvec k) u_{i, \alpha} Q_\alpha
\end{equation}
written as a matrix in the basis of localized Wannier orbitals, where $Q_\alpha$
denotes the displacement vector in the space of phonon modes $u_{i, \alpha}$.
Within the scope of this work, we restrict $\alpha$ to two of the three $T_{1u}$
ferroelectric modes. The second summand of $\hat H(\bvec k, \bvec Q)$ plays the
role of a displacement-induced potential acting on the electronic degrees of
freedom. Solving the Hamiltonian for a certain displacement $\bvec Q$ yields
displacement-dependent eigenvalues $\epsilon_b(\bvec k, \bvec Q)$. We
demonstrate that this model captures all relevant features of the DFT data
(cf.~\cref{fig:sto}~(c) of the main text) in \cref{fig:supp:model3}. We further
show Floquet spectral functions for both the valence- and conduction band sector
(projected to the Ti-$T_{2g}$ and the O-$p$ sectors) in
\cref{fig:supp:spectral}.

We note that we explicitly symmetrize the hopping parameters of the Wannier model using the $O_h$ point group. Furthermore, we simplify the electron-phonon coupling and only take into account the onsite and nearest-neighbor contributions for the three ferroelectric modes; those are enhanced by a factor larger than $10$. Also these terms are symmetrized.

\section{Aharonov-Anandan phases and the inverse Faraday effect}
In order to understand how the inverse Faraday effect
(IFE)~\cite{pitaevskii1961electric, vanderziel1965optically,
pershan1966theoretical, kirilyuk2010ultrafast, popova2011theory,
battiato2014quantum, juraschek2020phonomagnetic} can be seen as a limit of the
Aharonov-Anandan (AA) phase, we construct a model consisting of atomic
$p$-orbitals ($p_x$, $p_y$, $p_z$) and an $s$-orbital. By coupling these
orbitals to a harmonic potential that is displaced from the origin by $\bvec u$,
i.e., $V(\bvec r) = (\bvec r-\bvec u)^2/2$, we obtain the electron-nuclear
coupling to linear order:
\begin{equation}
    \braket{i|V(\bvec r)|j} = \mathrm{const.} - \braket{i|\bvec r\cdot\bvec u|j} + \mathcal{O}(\bvec u^2) = \mathrm{const.} + V_{ij} \,.
\end{equation}
As the $s$-orbital is an even function, and the $p$-orbtials are odd, the matrix
elements of the $\bvec r$ operator are nonzero only on the off-diagonal of the
orbital sectors. The electron-nuclear coupling written as matrix in the space of
$(s, p_x, p_y, p_z)$ orbtials reads
\begin{equation}
    V_{ij} \overset.= -g\begin{pmatrix}
        0 & u_x & u_y & u_z \\
        u_x & 0 & 0 & 0 \\
        u_y & 0 & 0 & 0 \\
        u_z & 0 & 0 & 0
    \end{pmatrix}\,.
\end{equation}
If we assume the nuclear coordinates to revolve around the origin in the
$xy$-plane, the $p_z$ orbital decouples from the system and it is sufficient to
consider an $s$- and the $p_{x,y}$-orbitals. Using $\bvec u(t) = u(\cos\Omega t,
\sin\Omega t, 0)^T$ as well as an on-site energy level $\Delta$ for the
$s$-orbital, we set up the Floquet Hamiltonian following \cref{app:floqexe}. The
Floquet hoppings read
\begin{equation}
    h_0 = \begin{pmatrix}
        \Delta & 0 & 0\\
        0 & 0 & 0\\
        0 & 0 & 0
    \end{pmatrix} \,, \qquad
    h_{\pm 1} = \frac{ug}2\begin{pmatrix}
        0 & 1 & \mp i \\
        1 & 0 & 0 \\
        \mp i & 0 & 0
    \end{pmatrix} \,,
\end{equation}
which are matrices in $(s,p_x,p_y)$ orbitals. In high-frequency approximation,
we set up an effective
Hamiltonian~\cite{rodriguez-vega2018floquet,oka2019floquet,vogl2020effective}
(see \cref{sec:supp:floqg}) and arrive at
\begin{equation}
    H_\mathrm{eff} = h_0 + h_{-1} \frac1{h_0-\Omega}h_1 + h_1\frac1{h_0+\Omega}h_{-1} \overset.= \begin{pmatrix}
        \Delta & 0 & 0 \\
        0 & 0 & 0 \\
        0 & 0 & 0
    \end{pmatrix} + \frac{\Delta u^2 g^2}{2\Delta^2 - 2\Omega^2} \begin{pmatrix}
        0 & 0 & 0 \\
        0 & 1 & 0 \\
        0 & 0 & 1
    \end{pmatrix} + \frac{\Omega u^2 g^2}{2\Delta^2 - 2\Omega^2} \begin{pmatrix}
        0 & 0 & 0 \\
        0 & 0 & i \\
        0 & -i& 0
    \end{pmatrix} \,.
    \label{eq:app:ife}
\end{equation}
The last term corresponds to an orbital angular momentum contribution, with the
same functional form as the one encountered in the inverse Faraday effect from
electron-phonon coupling~\cite{shabala2024phononinversefaradayeffect}.

The connection to the AA-phase~\cite{aharonov1987phase, oka2009photovoltaic,
oka2019floquet} can be made by constructing eigenstates of the Floquet
Hamiltonian in high-frequency approximation (as in \cref{app:floqexe}). First
order perturbation theory in $(\Delta\pm\Omega)^{-1}$
yields~\cite{rodriguez-vega2018floquet}
\begin{equation}
  \ket{p_\pm^{(1)}(t)} = \bigg(
  \frac{h_{+1} e^{-i\Omega t}}{h_0-\Omega} +
  \frac{h_{-1} e^{+i\Omega t}}{h_0+\Omega}
  \bigg) \ket{p_\pm} \,.
\end{equation}
The AA-phase $\gamma^\mathrm{AA}_\pm$ thus is given by
\begin{equation}
  \frac{\gamma^\mathrm{AA}_\pm}T = \frac1T \int_0^T\dd t \bra{p_\pm^{(1)}(t)}
  i\partial_t \ket{p_\pm^{(1)}(t)} = \dots = \Omega \bra{p_\pm} \bigg[
    \frac{h_{-1}}{h_0+\Omega}, \frac{h_{+1}}{h_0-\Omega} \bigg] \ket{p_\pm} =
  \pm\frac{\Omega u^2g^2}{2\Delta^2-2\Omega^2} \,.
\end{equation}
As $\gamma^\mathrm{AA}/T$ is the quantity entering the Hamiltonian, its
equivalence to the angular momentum term in \cref{eq:app:ife} demonstrates that
the phonon-IFE can be understood as a high-frequency limit of the
Aharonov-Anandan phase, i.e., electron-nuclear quantum geometry. Notably, the
formulation in terms of electron-nuclear quantum geometry (i.e. the AA-phase) is
more general than the phonon-IFE: While the IFE requires a model with
intermediate states for Raman processes, the quantum geometric picture can be
applied if those are not present as in the case of the $E\otimes e$ Jahn-Teller
model (see main text).

\section{Floquet optical conductivity}
\label{sec:supp:opcond}

The optical response function in Floquet theory can be written
as~\cite{oka2009photovoltaic, kumar2020linear, cupo2023optical}
\begin{equation}
    \label{eq:supp:sigma}
    \sigma_{\mu\nu}^{(n)} = \frac{i}{\omega} \sum_{\alpha} f_\alpha
    \dbra{\phi_{\alpha,0}} j_\mu\mathcal G(\omega + \epsilon_\alpha + n\Omega)
    j_\nu + j_\nu\mathcal G^\dagger(-\omega + \epsilon_\alpha) j_\mu
    \dket{\phi_{\alpha,n}} \,,
\end{equation}
where the index $n$ corresponds to responses at frequency $\omega+n\Omega$ and
we omit the diamagnetic contribution. We use the following notation for the
Floquet expectation value of an operator $\mathcal O$:
\begin{equation}
    \dbra{\phi_{\alpha,n}} \mathcal O \dket{\phi_{\beta,m}} =
    \frac1T\int_0^T\!\!\dd t\,\braket{\phi_{\alpha,n}(t) | \mathcal O |
    \phi_{\beta,m}(t) }\,,
\end{equation}
with $\ket{\phi_{\alpha,n}(t)}$ being the $n$-th Floquet block of the Floquet
state $\alpha$. We here restrict ourselves to considering the $n=0$ component of
$\sigma^{(n)}$ [\cref{eq:supp:sigma}], i.e. the so-called \emph{homodyne}
component.

\Cref{eq:supp:sigma} contains occupations of Floquet states $f_\alpha$. We aim to find those Floquet states that can be adiabatically connected to the valence band manifold of the equilibrium system. Note that ``adiabaticity''{} is well-defined in this case: The driving frequency $\Omega$ as well as the ferroelectric amplitude/electron-phonon coupling $gQ$ fulfill $\Omega \ll \Delta$ and $gQ \ll \Delta$, where $\Delta$ is the band gap. To find the correct states, we employ the following procedure:
\begin{enumerate}
  \item Create a mask, and mask all states that (i)~have a Floquet eigenvalue
    $\epsilon^i \geq 0$ (with the chemical potential adjusted such that $0$ lies
    in the center of the valence/conduction band gap) or (ii)~an eigenvalue
    index $i \geq N\cdot 9/12$ (with eigenvalues sorted by $\epsilon^i$ in an
    ascending manner, such that states in the valence band manifold are
    selected).
  \item \label{supp:iterlabel} Find the state $\dket{\psi_i}$ (among the ones
    that are not masked) that has maximum weight in the $n=0$ sector, smeared
    with a Gaussian of width $\sigma_\mathrm{floq}=3\Omega$ in Floquet replica
    space. Append the index $i$ of this state to the list of adiabatically
    connected valence band states
  \item Iterate over the $n$ boosted states $\mathcal B_n \dket{\psi_i}$ and
    mask those eigenstates with maximum overlap to each of the boosted ones. The
    boost operator $\mathcal B_n$ is defined as
    \begin{equation}
      \mathcal B_n \dket{\psi_{\alpha,m}} = \dket{\psi_{\alpha,m+n}} \,,
    \end{equation}
    i.e., it maps a Floquet state $\dket{\psi_{\alpha,m}}$ to its replica
    displaced by $n\Omega$.
  \item If the length of the list of adiabatically connected valence band states
    is $9$, all valence band states are found. Otherwise, go to
    \cref{supp:iterlabel} and continue from there.
\end{enumerate}
Having obtained a list of $9$ adiabatically connected eigenstates that represent
the valence band manifold (the number ($9$) $12$ merely serves as counter of the
number of (valence) bands in the equilibrium model), we set the occupations of
these states to $f_\alpha=1$, and the occupations of all other states to zero.
The such obtained valence band manifold of the driven model is the one that is
adiabatically connected to the equilibrium valence band manifold upon increasing
the electron-phonon coupling $gQ$.

\subsection{Floquet tight-binding current operators}
To obtain the current operators $j_\mu$ in the Wannier basis, we make use of the
fact that the Floquet Hamiltonian has nontrivial blocks only on the main and
first neighboring diagonal, i.e., $h_0$ and $h_{\pm1}$
(cf.~\cref{eq:wannierfloq} of the main text). By means of discrete Fourier
transformation, we obtain their real-space representations:
\begin{equation}
    \hat{h}_0(i,j) = \sum_{\bvec k} e^{i\bvec k(\bvec R_i-\bvec R_j)} \hat H_W(\bvec k) \,, \quad
    \hat{h}_{\pm1}(i,j) = \sum_{\bvec k} e^{i\bvec k(\bvec R_i-\bvec R_j)} \sum_l \frac Q2 \hat g_l(\bvec k)\big(u_{l,x}\pm iu_{l,y}\big) \,,
\end{equation}
with $\hat H_W(\bvec k)$ the Wannier Hamiltonian in momentum space, $\bvec R_i$
lattice vectors and $\hat g_l(\bvec k)$ the electron-phonon coupling
corresponding to phonon modes $u_{l, \alpha}$ with $\alpha\in\{x,y\}$. We find
the Floquet components of the current operators as distance/direction modulated
hopping parameters:
\begin{equation}
    \bvec j_n^{\alpha,\beta}(i,j) = \big( \bvec R_i - \bvec R_j + \bvec r_\alpha - \bvec r_\beta \big)\, h_n^{\alpha,\beta}(i,j) \,,
\end{equation}
where $\alpha,\beta$ correspond to orbital indices, with $\bvec
r_{\alpha,\beta}$ denoting their position in the unit cell. The Floquet block is
indexed by $n\in\{0,\pm1\}$. Fourier transformation yields the momentum space
representation $\hat{\bvec j}_n(\bvec k)$
\begin{equation}
    \hat{\bvec j}_n(\bvec k) = \sum_{\bvec R = \bvec R_i - \bvec R_j} e^{-i\bvec k\bvec R} \hat{\bvec j}_n(i,j) \,.
\end{equation}
The full Floquet space current operator, $\bvec j$, can be assembled in the same
manner as the Floquet Hamiltonian, i.e.,
\begin{equation}
    \hat{\bvec j}(\bvec k) = \begin{pmatrix}
        \ddots & \hat{\bvec j}_1(\bvec k) \\
        \hat{\bvec j}_{-1}(\bvec k) & \hat{\bvec j}_0(\bvec k) & \hat{\bvec j}_1(\bvec k) \\
        & \hat{\bvec j}_{-1}(\bvec k) & \hat{\bvec j}_0(\bvec k) & \hat{\bvec j}_1(\bvec k) \\
        & & \hat{\bvec j}_{-1}(\bvec k) & \hat{\bvec j}_0(\bvec k) & \hat{\bvec j}_1(\bvec k) \\
        & & & \hat{\bvec j}_{-1}(\bvec k) & \ddots
    \end{pmatrix} \,.
\end{equation}
For brevity, \cref{eq:supp:sigma} implicitly accounts for the matrix structure
of the orbitals as well as momentum, i.e., $\bvec j \equiv \hat{\bvec j}(\bvec
k)$. The summation over $\alpha$ is understood as summation over bands and
momenta. In addition, \cref{eq:supp:sigma} references cartesian components of
$\bvec j = (j_x, j_y, j_z)^T$ using indices $\mu,\nu$.

\subsection{Technical details}
\begin{figure}
    \centering
    \includegraphics{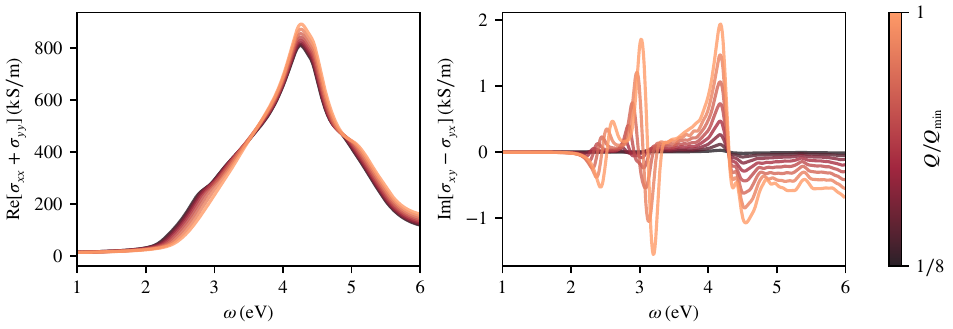}
    \includegraphics{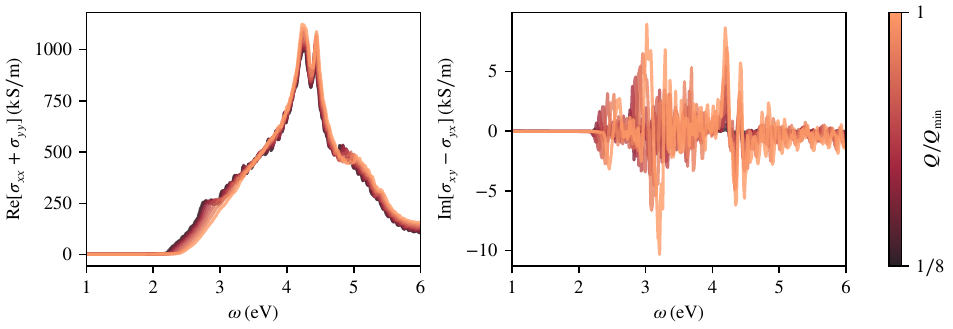}
    \caption{Real part of the homodyne Floquet optical conductivity
    $\sigma^{(0)}_{\mu\nu}(\omega)$ for \STO{}. We show the diagonal and Hall
    components in the left and right panels, respectively. The color encodes the
    ferroelectric amplitude $Q$ relative to the ferroelectric minimum
    $Q_\mathrm{min}$. The lower row corresponds to a strongly reduced broadening $\eta=0.01\,\mathrm{eV}$, which serves the purpose of (i)~demonstrating that the Hermitian components are strictly zero inside the optical gap and (ii)~the results for $\eta=0.1\,\mathrm{eV}$ (upper row) are well converged.}
    \label{fig:supp:sigma}
\end{figure}

We evaluate \cref{eq:supp:sigma} using a momentum grid of $N_{\bvec k}=64^3$
points in the BZ and with a frequency broadening of $\omega \leftarrow \omega +
i\eta = \omega + i\,0.1\,\mathrm{eV}$. The drive frequency is set to
$\Omega=0.03\,\mathrm{eV}\approx 7\,\mathrm{THz}$, and we truncate the Floquet
expansion after reaching $\pm60\Omega$. Such large values are required for
converging the procedure that finds those states adiabatically connecting the
driven to the equilibrium valence band manifold (see the beginning of this
section). In addition to the large cutoff for Floquet replica, we must break
point-group symmetries by an infinitesimal amount in order to lift numerical
degeneracies in a consistent manner across different Floquet replica.
\Cref{fig:supp:sigma} shows the resulting optical conductivity for several
amplitudes $Q$. We note that the band gap is underestimated due to the
well-known band-gap problem in semilocal density functional theory
(cf.~\cref{fig:supp:model3}), resulting in a DFT band-gap of
$\Delta\approx2.3\,\mathrm{eV}$ vs. an experimental band-gap of more than
$3\,\mathrm{eV}$.

\subsection{Faraday and Kerr rotation}

\begin{figure}
    \centering
    \includegraphics{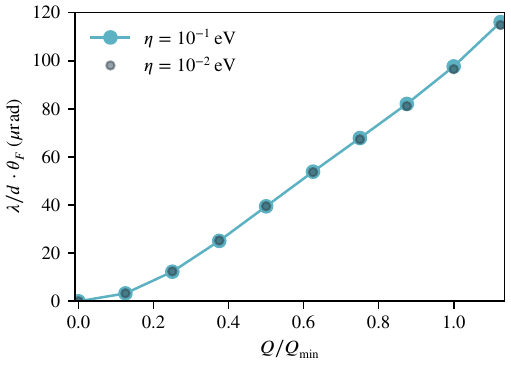}
    \caption{Faraday rotation $\theta_F$ as a function of ferroelectric amplitude $Q/Q_\mathrm{min}$ for probe frequencies deep inside the optical gap ($\omega=1.1\,\mathrm{eV}$). The Faraday rotation $\theta_F$ is proportional to the ratio of penetration depth over wavelength, i.e., $\theta_F\propto d/\lambda$. We multiply out this ratio on the $y$-axis and plot $\lambda \theta_F/d$.}
    \label{fig:supp:faraday}
\end{figure}

We calculate the Faraday rotation $\theta_F$ following Ref.~\cite{brynildsen2013verdet}:
\begin{equation}
    \theta_F = \frac{\pi d (\eta_- - \eta_+)}{\lambda} \,,
\end{equation}
where $d$ denotes the sample thickness, $\lambda$ the wavelength, and $\eta_\pm$ the real part of the refractive index for left/right circularly polarized light. The refractive index $n = \eta + i \kappa$ is obtained from the dielectric function $\epsilon$ which is related to the optical conductivity:
\begin{align}
    \epsilon_{ij} &{}= \matrixOne_{ij} + \frac{i \sigma_{ij}}{\hbar\omega\epsilon_0} \,, \\
    n_\pm &{}= \sqrt{ \epsilon_{xx} \pm i \epsilon_{xy} } \,.
\end{align}
\Cref{fig:supp:faraday} displays the Faraday rotation at
$\omega=1.1\,\mathrm{eV}$ as a function of ferroelectric amplitude $Q$---due to the well-known band gap problem in semi-local DFT, the band gap in our model is $\Delta\approx2.3\,\mathrm{eV}$ as opposed to the real material, where $\Delta_{\mathrm{\STO}}\approx3.6\,\mathrm{eV}$, so we discuss $\omega=1.1\,\mathrm{eV}$ as in-gap frequency. For
small $Q$, a quadratic behavior is clearly visible. Considering that the
experiment by Basini~\emph{et.al.}~\cite{basini2022terahertz} estimates the
penetration depth of the Terahertz pump pulse to roughly four wavelengths of the
probe pulse, our result for the Faraday rotation is well in line with their
results: $\theta_F \lesssim 200\,\mathrm{\mu rad}$.

In a half-infinite system, the real part of the Kerr angle is zero without
dissipation, i.e., inside the optical gap.~\cite{suits1972faraday,
you1998generalized, mansuripur2000magneto, protopopov2014practical}. The
experimental setup in Ref.~\cite{basini2022terahertz} might correspond to a
magnetic thin-film rather than a semi-infinite slab. Therefore, Fabry-Perot-like
repeated scattering at the top and bottom surfaces can lead to a nonzero Kerr
rotation that is largely determined by the Faraday rotation. Indeed, the surface
magneto-optical Kerr effect (SMOKE)~\cite{qiu2000surface} or Kerr rotation for
2D systems~\cite{tse2010giant, catarina2020magneto} allow for a nonzero real
part of the Kerr rotation without dissipative components of the optical
conductivity. In addition, Basini~\emph{et.al.}~\cite{basini2022terahertz}
observe a Kerr rotation that is comparable to the value of the Faraday rotation,
which would be in line with an interpretation as SMOKE or thin-film effect.

\vfill
\makesuppbib
\supplementEnd

\end{document}